\documentclass[letterpaper, 10 pt, conference]{ieeeconf}  

\IEEEoverridecommandlockouts                              

\usepackage{graphicx,epstopdf} 
\usepackage{amsmath,amssymb,amsfonts}
\usepackage{multirow}
\usepackage{environ}         
\usepackage{etoolbox}        
\usepackage{textcomp}
\usepackage{algorithm,algpseudocode}
\usepackage{xcolor}
\usepackage{tikz}
\usepackage{mathtools}
\usepackage{setspace}
\usepackage{array}
\usepackage{calc}
\usepackage{makecell}
\usepackage{tabularx}
\usepackage{booktabs}
\usepackage{makecell}
\usepackage{pgfplots,pgfplotstable}
\pgfplotsset{compat=1.16}
\usepgfplotslibrary{statistics}
\usepackage{scrextend}
\usepackage{tabularray}
\usepackage{siunitx}
\usepackage{caption}
\usepackage{subcaption}
\usepackage{cite}
\usepackage{fancyhdr}
\usepackage[colorlinks,linkcolor=blue,citecolor=red,urlcolor=blue,bookmarks=false,hypertexnames=true]{hyperref} 
\makeatletter
\newcommand*{\rom}[1]{\expandafter\@slowromancap\romannumeral #1@}
\makeatother

\newtheorem{definition}{Definition}

\title{\LARGE \bf
Lateral String Stability in Autonomous \& Connected Vehicle Platoons}

\author{%
    Neelkamal Somisetty and Swaroop  Darbha$^{*}$ \\[1ex]
    \small Department of Mechanical Engineering, Texas A\&M University, College Station, TX, USA\\[1ex]
    \small Email: \{neelkamal.sept18, dswaroop\}@tamu.edu
}
\begin{document}

\maketitle
\thispagestyle{plain}
\fancypagestyle{plain}{
  \fancyhf{}
  \fancyfoot[C]{\footnotesize 979-8-3503-8727-8/24/\$31.00 \copyright2024 IEEE}
  \renewcommand{\headrulewidth}{0pt}
  \renewcommand{\footrulewidth}{0pt}
}

\begin{abstract}
This paper addresses the lateral control of Autonomous and Connected Vehicles (ACVs) in a platoon executing an Emergency Lane Change (ELC) maneuver. These maneuvers are typically triggered by emergency signals from the front or rear of the platoon in response to the need to avoid obstacles or allow other vehicles to pass. The study assumes that ACVs maintain reliable connectivity, enabling each following vehicle to access GPS position traces of both the lead and immediately preceding vehicles in the platoon. We demonstrate that lateral string stability in the ACV platoon can be achieved using communicated information solely from the lead and preceding vehicles. Additionally, we present a lateral control framework for ACVs, which helps track a discretized preview of the trajectory constructed from the communicated data. This framework involves constructing two distinct trajectories based on the preview data from the lead and preceding vehicles, calculating the associated errors and lateral control actions for each, and then integrating these to generate a steering command. Numerical results validate the effectiveness of the proposed lateral control scheme.
\end{abstract}

\section{INTRODUCTION}

Maintaining a desired and safe trajectory is essential for the lateral control of autonomous vehicles. Numerous lateral controllers have been proposed in the literature for lane-keeping, which involves keeping the vehicle within lane boundaries \cite{lee2018synthesis, marino2011nested}. This study focuses on designing a lateral controller for Autonomous and Connected Vehicles (ACVs) in a platoon during an Emergency Lane Change (ELC), which is also suitable for lane-keeping. ELC maneuvers may be initiated by an obstacle avoidance action from the front or by an approaching emergency vehicle from behind. For such scenarios, the lateral controllers proposed for ACVs in the literature often rely on road infrastructure elements like reference wires, magnets, embedded signals, or lane markers \cite{alleleijn2014lateral, tan2003automatic, taylor1999comparative}. This dependency presents challenges related to cost and vulnerability to adverse weather conditions such as snow or smog. Moreover, in situations requiring close following distances, like eco-driving, or on roads without lane markings, detecting these markings may not be feasible \cite{liu2020lateral_ai}. Research on  vehicle following for lane-free driving environments underscores the difficulties and solutions for operating without traditional road infrastructure \cite{theodosis2023cruise}. Consequently, there is a need for developing lateral control schemes that operate independently of road infrastructure, leveraging the advantages of vehicle connectivity instead.

\begin{figure}[ht]
\begin{center}
  \includegraphics[scale=0.4]{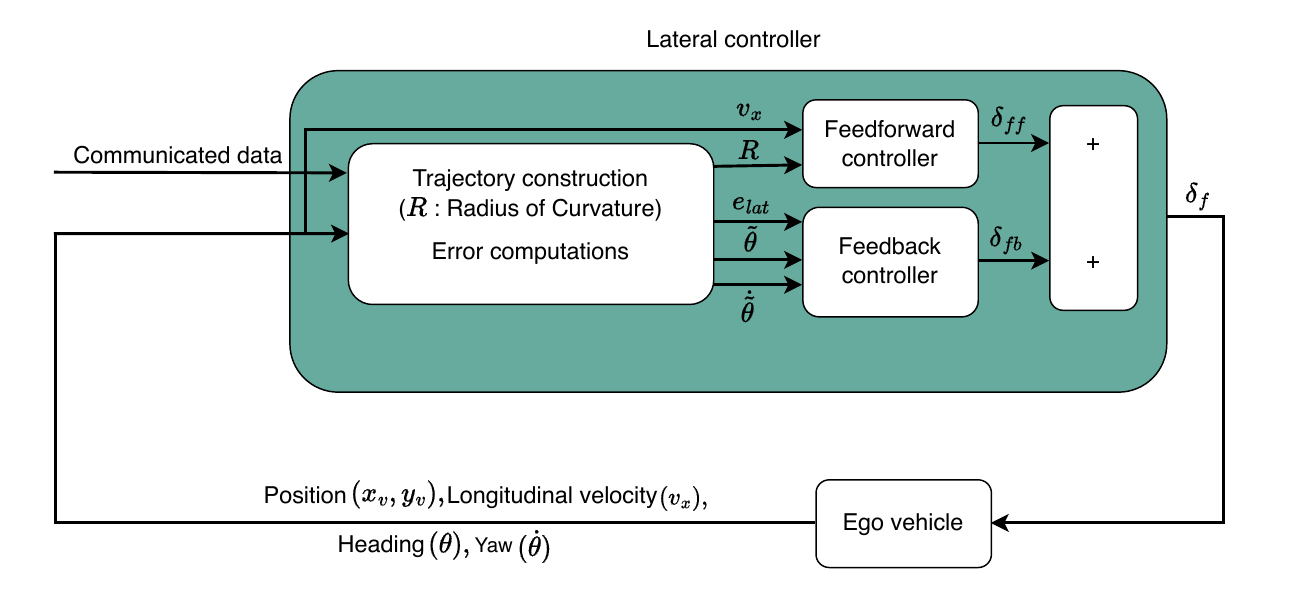}  
\end{center}
\caption{Structure of the proposed lateral controller.}
\label{fig:ControllerStructure}
\end{figure}

This study addresses the challenge of controlling the lateral dynamics of an ACV in a platoon with limited preview information from the lead and preceding vehicles. During an ELC maneuver, the lead vehicle performs a lane change to avoid the obstacle, constructs the trajectory in real-time, and communicates this information to its followers. It is assumed that the maneuver is not extreme, as the lead vehicle has the best view or information from infrastructure about the obstacle and does not need to follow any vehicle ahead. The proposed lateral controller typically consists of both feedforward and feedback components as shown in Fig. \ref{fig:ControllerStructure}.

The feedforward controller uses the preview information to construct a trajectory to be tracked, based on which a feedforward steering angle input is computed. If the lateral vehicle dynamic model were known exactly, and the vehicle were following the desired trajectory with no lateral and heading errors, such an input would maintain the vehicle on the trajectory in the absence of disturbances. However, since this ideal scenario is rarely encountered in practice, a feedback controller is also necessary. The task of a feedback controller is to get the vehicle back onto the desired trajectory in the presence of disturbances, model uncertainties, and initial errors in heading and lateral position and their rates. It relies on the cross-track error \(e_{lat}\) (the distance between the ACV and the closest point on the target trajectory), heading error \(\tilde{\theta}\) (the deviation between the ACV’s heading and the direction of the tangent to the target trajectory at the closest point), and yaw rate error \(\dot{\tilde{\theta}}\). Within this framework, \textit{two target trajectories} are constructed based on the information communicated from the lead and immediate preceding vehicles, respectively. Connectivity is crucial for executing the proposed lateral control framework, as it enables each ACV in the platoon to access the position information of both the lead ACV and its immediate preceding ACV. 

\subsection{Literature Review and Novel Contributions}
A concise overview of lane-changing maneuvers and associated control methods is provided in previous studies \cite{badnava2021platoon, bevly2016lane}. In contrast to methods that employ vision systems, magnetometers, or guided wire technology \cite{taylor1999comparative, lu2004practical, fenton1976steering}, the proposed approach exclusively utilizes communicated positional data from both the lead and preceding vehicles, along with GPS/IMU measurements to determine the ego vehicle’s position, heading, yaw rate, and longitudinal velocity (as illustrated in the outputs of the vehicle block in Fig. \ref{fig:ControllerStructure}).

The proposed lateral controller framework includes both feedforward and feedback components. Such a framework has been previously used in various studies \cite{fenton1976steering, kapania2015design, liu2019lateral, liu2020lateral, liu2024lateral}. This work differs from \cite{kapania2015design} as the following ACVs do not have a pre-planned trajectory to track; instead, trajectories are constructed in real-time based on the limited GPS measurements communicated. Unlike the works in \cite{liu2019lateral, liu2020lateral}, we propose a lateral control scheme for a platoon with more than two ACVs. Additionally, the method we employ to utilize the communicated data from the lead and preceding vehicles for enhanced performance differs from our previous work in \cite{liu2024lateral}.

In our prior work \cite{liu2024lateral}, we demonstrated lateral string instability using predecessor-only information, even when the lead vehicle was tracking a straight line. In contrast, this study provides a theoretical proof establishing the lateral string stability of a platoon utilizing communicated information from both lead and preceding vehicles while executing a straight-line maneuver. This theoretical proof complements the numerical simulations presented in the dissertation of Liu \cite{liu2020dissertation}, which are illustrated in this paper to offer comprehensive validation of the proposed control scheme.

The novel contributions of this paper are significant to the field of autonomous vehicles and are twofold. Firstly, we demonstrate the lateral string stability of a platoon utilizing communicated information from both lead and preceding vehicles while executing a straight-line maneuver. Motivated by this finding, we propose a lateral control framework designed for platoons with more than two ACVs, expanding its applicability beyond previous works. A key innovation in our framework is the method we employ to utilize communicated data from both the lead and preceding vehicles, enhancing performance and scalability for larger platoons.

\subsection{Organization}

The structure of this paper is as follows: Section \ref{sec:model} provides a brief overview of the lateral dynamic model of the vehicle. Section \ref{sec:design} introduces the proposed lateral control scheme. Section \ref{sec:stability} presents the proof for the lateral string stability of the ACV platoon using the lead and preceding vehicles information. Section \ref{sec:results} presents the simulation results. Finally, Section \ref{sec:conclusions} concludes the paper with final remarks.

\section{LATERAL DYNAMIC MODEL of an ACV}
\label{sec:model}

The widely adopted “bicycle” model serves as the fundamental framework for lateral vehicle control \cite{karafyllis2022constructing, liu2019lateral, hassanain2020string}. Here, let \(m_v\) denote the vehicle’s mass and \(I_z\) its moment of inertia about the center of mass. The parameters \(C_f\) and \(C_r\) represent the cornering stiffness values for the front and rear axles, respectively, while \(a\) and \(b\) denote the distances from the center of gravity to the front and rear axles. For a vehicle with front-wheel steering, given a steering angle input \(\delta_f\), Newton-Euler equations combined with linear constitutive models for tire cornering forces result in the following dynamic equations:

\begin{eqnarray}
m_v(\frac{dv_y}{dt}+v_x\dot{\theta})=C_f\delta_f-\frac{C_f+C_r}{v_x}v_y-\frac{aC_f-bC_r}{v_x}\dot{\theta},\label{ForceEOM}\\
    I_z\ddot{\theta}=aC_f\delta_f - \frac{aC_f-bC_r}{v_x}v_y-\frac{a^2C_f+b^2C_r}{v_x}\dot{\theta}.\label{MomentEOM}
\end{eqnarray}


To facilitate the design of a feedback controller, it is advantageous to reformulate the equations of motion in terms of position and heading deviations from the desired trajectory. Following previous studies \cite{liu2019lateral, liu2020lateral}, the vehicle dynamic equations (\ref{ForceEOM}) and (\ref{MomentEOM}) can be represented using the state variables \(e_{lat}, \dot{e}_{lat}, \tilde{\theta}, \frac{d\tilde{\theta}}{dt}\) in the form of the following matrices:
\begin{eqnarray}
\mathbf{M} &:= &\begin{bmatrix}
m_v & 0 \\
0 & I_z
\end{bmatrix}, \quad
\mathbf{C} \ := \begin{bmatrix}
\frac{C_f + C_r}{V_0} & \frac{aC_f - bC_r}{V_0} \\
\frac{aC_f - bC_r}{V_0} & \frac{a^2C_f + b^2C_r}{V_0}
\end{bmatrix},\nonumber \\
\mathbf{B} &:= &\begin{bmatrix}
1 \\
a 
\end{bmatrix}, \quad
\mathbf{F} \ := \begin{bmatrix}
m_vV_0^2 + (aC_f - bC_r) \\
a^2C_f + b^2C_r
\end{bmatrix},\nonumber \\
\mathbf{L} &:= &\begin{bmatrix}
0 & -(C_f + C_r) \\
0 & -(aC_f - bC_r)
\end{bmatrix}, \quad 
\mathbf{x} \ := \begin{bmatrix}
e_{lat} \\
\tilde \theta
\end{bmatrix}.\nonumber 
\end{eqnarray}

Here, we assume the vehicles longitudinal speed is a constant, say $v_x = V_0$. The governing equations can thus be written as:
\begin{eqnarray}
\mathbf{M} \ddot{\mathbf{x}} + \mathbf{C} \dot{\mathbf{x}} + \mathbf{L} \mathbf{x} = \mathbf{B} C_f \delta_f - \mathbf{F} \left(\frac{1}{R}\right).
\label{eq:System}
\end{eqnarray}

Building on our earlier work \cite{liu2020lateral}, we employ a second-order model for steering actuation dynamics that has been experimentally validated, analogous to the model utilized in \cite{hassanain2020string}. The transfer function for actuation is represented as follows:

\begin{align}
       H_a(s) = \frac{w_n^2}{s^2 + 2\zeta w_ns + w_n^2},
       \label{Eq:Transfer_fn}
\end{align}
where
\begin{align*}
    w_n = \sqrt{\frac{K_r}{J_w}}, \quad \zeta = \frac{b_w}{\sqrt{K_rJ_w}}.
\end{align*}

Here, \(\zeta\) and \(\omega_n\) are the damping ratio and the natural frequency of steering actuation, respectively. \(J_w\) represents the steering wheel inertia, \(b_w\) is the torsional viscous damping coefficient, and \(K_r\) is the torsional stiffness of the steering column.

\section{Lateral Controller Design}
\label{sec:design}

This section outlines the design of the lateral controller, depicted in Fig. \ref{fig:ControllerStructure}. The design process involves the following key tasks:
1. Construction of the target trajectory and computation of feedback error signals relative to the target trajectory. 2. Design of the feedforward and feedback controllers. The former task is discussed in detail in our prior work \cite{liu2024lateral}. However, we present an overview here for the sake of completeness.

\subsection{Target Trajectory Construction and Feedback Error Signals Computation}

To enable effective lateral control, this subsection presents a method for constructing the target trajectory using preview data from lead and preceding vehicles in the platoon, along with computing feedback error signals relative to this trajectory. The trajectory is built using circular arc spline approximations of sampled positional data from lead and preceding ACVs, with adjustments based on preview distance. This approach simplifies the computation of curvature, cross-track, heading, and yaw rate errors by utilizing commonly structured road geometries.

\begin{figure}[ht]
    \centering
    \includegraphics[scale=0.6]{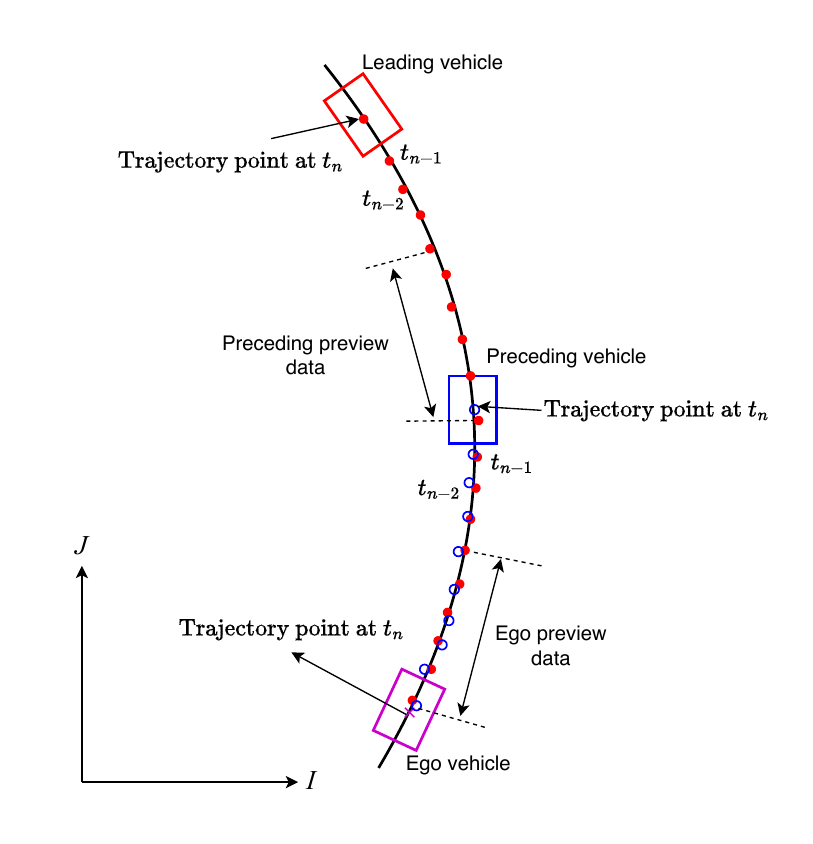}
    \caption{Sampled trajectories of ACVs in the platoon and information accessible to the ego ACV.}
    \label{fig:leadandfollower}
\end{figure}

As illustrated in Fig. \ref{fig:leadandfollower}, the ego ACV collects position data within a preview distance, \(L_{preview}\), and processes it to construct two separate ELC trajectories. For trajectory construction, points are sequentially evaluated to identify segments as either straight lines or circular arcs, based on a threshold error metric that ensures accurate representation within the previewed path. The feedback error signals—lateral error (\(e_{lat}\)), heading error (\(\tilde{\theta}\)), and yaw rate error (\(\dot{\tilde{\theta}}\))—are derived from the trajectory, enabling effective tracking by the lateral controller.

\subsection{Lateral Controller Synthesis}
The lateral controller for tracking a trajectory can be decomposed into two parts: a feedforward component and a feedback component, i.e.,
\begin{equation}
    \delta_c = \delta_{ff} + \delta_{fb}.
\end{equation}

\subsubsection{\bf Feedforward controller design}
The feedforward controller provides the steering input required to keep an ACV on a circular trajectory without feedback, assuming ideal initial conditions and the absence of disturbances. The feedforward steering input, \(\delta_{ff}\), is the input required for the ACV to follow a circular trajectory with radius \(R\) while maintaining a longitudinal speed of \(V_0\). The feedforward steering input is given by:
\begin{equation}
    \delta_{ff} = \frac{a+b}{R} + K_{sg} \frac{V_0^2}{R},
\end{equation}
where \(K_{sg}\) is the steer gradient. Let \(\alpha \in [0,1]\) represent the weighting factor for the preview data from the preceding ACV with respect to tracking. The feedforward steering input is then defined by:
\begin{eqnarray}
   \nonumber \delta_{ff} &=& \alpha \left(\frac{a+b}{R_p} + K_{sg} \frac{V_0^2}{R_p}\right)\\ && + (1-\alpha) \left(\frac{a+b}{R_l} + K_{sg} \frac{V_0^2}{R_l}\right), 
\end{eqnarray}

where \(R_l\) and \(R_p\) are the radii of the trajectory calculated using the preview data from the lead and preceding ACVs, respectively.

\subsubsection{\bf Feedback Controller Design}

The feedback control strategy utilizes the available error signals, specifically \(e_{lat}\), \(\tilde{\theta}\), and \(\dot{\tilde{\theta}}\), relative to the ELC trajectory. Let \(e_{lat,l}\), \(\tilde{\theta}_l\), and \(\dot{\tilde{\theta}}_l\) denote the cross-track error, heading error, and yaw rate error, respectively, for the ELC trajectory derived from the lead ACV’s data. Similarly, let \(e_{lat,p}\), \(\tilde{\theta}_p\), and \(\dot{\tilde{\theta}}_p\) represent the corresponding errors for the ELC trajectory based on the preview data from the preceding ACV. 

In the absence of lateral velocity information, we aim to design a control law in the following form:
\begin{eqnarray}
   \nonumber \delta_{fb} = &-&(1-\alpha) (k_e e_{lat,l} + k_{\theta} \tilde{\theta}_l + k_{\omega} \dot{\tilde{\theta}}_l)\\ &-& \alpha (k_e e_{lat,p} + k_{\theta} \tilde{\theta}_p + k_{\omega} \dot{\tilde{\theta}}_p), 
\end{eqnarray}
where the gains \(k_e\), \(k_{\theta}\), and \(k_{\omega}\) need to be determined.

\subsubsection{\bf Constructing the set of stabilizing structured feedback controllers}
The selection of gains \(k_e\), \(k_{\theta}\), and \(k_{\omega}\) is crucial for the implementation of the lateral control system. Various methods exist to construct the set of controllers \cite{malik2008linear, bhattacharyya2018linear, henrion2003positive} in the parameter space. In this approach, we utilize the D-decomposition technique from \cite{bhattacharyya2018linear} to derive the set of stabilizing feedback gains.

Applying the Laplace transform to both sides of equation \eqref{eq:System} and denoting \(\mathbf{x}_i(s)\) as the Laplace transform of \(\mathbf{x}_i(t)\), we arrive at:

\begin{equation}
    \mathbf{M}_0(s) \mathbf{x}(s) = \mathbf{B} C_f \delta_{f}(s).
\end{equation}

Given that \(\delta_f(s) = H_a(s) \delta_c(s)\) and accounting for the second-order steering actuation model described in \eqref{Eq:Transfer_fn}, the characteristic polynomial can be formulated in terms of the control parameter vector \(\mathbf{K} = (k_e, k_{\theta}, k_{\omega})\) as follows:
\begin{eqnarray}
\nonumber
\Delta(s; \mathbf{K}) \hspace{-0.75em}&=& \hspace{-0.75em}(s^2 + 2\zeta\omega_n s + \omega_n^2) \Delta_o(s) \\
\nonumber
&&\hspace{-0.75em}+ k_e C_f \omega_n^2 \left( I_z s^2 + \frac{b(a + b)C_r}{V_0} s + (a + b) C_r \right) \\
&&\hspace{-0.75em}+ (k_{\theta} + k_{\omega} s) C_f \omega_n^2 \left( m_v a s^2 + \frac{(a + b)C_r}{V_0} s \right),
\label{eq:cl_poly}
\end{eqnarray}
where
\begin{eqnarray*}
    \Delta_o(s) \hspace{-0.75em}&=& \hspace{-0.75em}s^2 \left( m_v I_z s^2 + \frac{(I_z + m_va^2)C_f + (I_z + m_vb^2)C_r}{V_0} s \right. \\
&& \hspace{-0.5em}\left. + \frac{(a + b)^2 C_f C_r}{V_0^2} - m_v(a C_f - b C_r) \right).
\end{eqnarray*}
Please refer to \cite{liu2020lateral} for a detailed derivation of equation \eqref{eq:cl_poly}.

A fundamental problem regarding the stability of tracking is to determine the set of control gains, \(\mathbf{K}\), such that the closed-loop characteristic polynomial in equation (\ref{eq:cl_poly}) is Hurwitz. One advantage of identifying the complete set of stabilizing gains is that it can be refined to satisfy additional performance requirements.

The D-decomposition approach leverages the continuous
dependence of the roots of a polynomial on its coefficients
for any regular perturbation. This method involves dividing the parameter space into distinct, signature-invariant regions by determining their boundaries. These boundaries are identified by (a) finding the values of \(\mathbf{K}\) such that \(\Delta(0, \mathbf{K}) = 0\), and (b) finding the values of \(\mathbf{K}\) for which \(\Delta(jw, \mathbf{K}) = 0\) for some \(w\). Once the parameter space is segmented, each region can be sampled to identify which sections correspond to a Hurwitz \(\Delta(s; \mathbf{K})\).

It is essential to recognize that the set of stabilizing controllers depends on the specific operating longitudinal speed \(V_0\), with \(\Delta(s; \mathbf{K})\) having a quadratic dependence on \(\frac{1}{V_0}\). By defining the parameter \(\gamma := \frac{1}{V_0}\), we can then construct the set of stabilizing controllers for each \(\gamma \in \left[\frac{1}{V_{max}}, \frac{1}{V_{min}}\right]\), where \(V_{min}\) and \(V_{max}\) represent the lower and upper limits of operating longitudinal speeds. A control gain vector \(\mathbf{K} = (k_e, k_{\theta}, k_{\omega})\) is then selected from the intersection of stabilizing sets as \(\gamma\) varies from \(\frac{1}{V_{max}}\) to \(\frac{1}{V_{min}}\).

\section{LATERAL STRING STABILITY for a PLATOON}
\label{sec:stability}

In platooning applications, lateral control must address two key issues: (a) lateral stability and (b) lateral string stability. While lateral stability has been discussed in the previous section, this section demonstrates the lateral string stability of a platoon executing a straight-line maneuver, i.e., when \(R = \infty\). Let ${\bf x}_i(t)$ denote the error in the state of the $i^{th}$ following vehicle in the convoy at time $t$ in tracking the trajectory. Essentially, the components of ${\bf x}_i (t)$ are $e_{lat}(t)$ with respect to the straight line trajectory and $\tilde \theta (t)$  as defined in the earlier subsection. We will define a weak definition of lateral string stability below.

\begin{definition}[\textbf{Lateral String Stability}] A convoy of vehicles is laterally string stable, if given an $M >0$,  independent of the size of the convoy, such that
$$\|{\bf x}_i(t) \| \le M \|{\bf x}_{1}(t)\|.$$
\end{definition}
\noindent {\bf Remark:} Essentially, small errors in tracking do not amplify with the size of the convoy thereby enabling the state errors to remain bounded throughout.

With instantaneous actuation, the governing equation in \eqref{eq:System} for the \(i^{th}\) vehicle is:
\begin{equation}
    \mathbf{M} \ddot{\mathbf{x}}_i + \mathbf{C} \dot{\mathbf{x}}_i + \mathbf{L} \mathbf{x}_i = \mathbf{B} C_f \delta_{f,i}.
\end{equation}

Taking the Laplace transform on both sides and letting \(\mathbf{x}_i(s)\) denote the Laplace transform of \(\mathbf{x}_i(t)\), we obtain:

\begin{equation}
 \mathbf{M}_0(s) \mathbf{x}_i(s) = \mathbf{B} C_f \delta_{f,i}(s),
\end{equation}
where 
\begin{eqnarray*}
&& \hspace{-1em} \mathbf{M}_0(s) := \\
&& \hspace{-1em} \begin{pmatrix}
        m_vs^2 + \frac{C_f + C_r}{V_0}s & \frac{aC_f - bC_r}{V_0}s - (C_f + C_r) \\
        \frac{aC_f - bC_r}{V_0}s & I_zs^2 + \frac{a^2C_f + b^2C_r}{V_0}s - (aC_f - bC_r)
    \end{pmatrix}.
\end{eqnarray*}

Considering the information from the lead and preceding vehicles, the control law is given by:
\begin{eqnarray}
  \nonumber \delta_{f,i} (s) = \hspace{-1em} &&-\begin{pmatrix} k_e & k_{\theta} + k_w s \end{pmatrix} \cdot\\
        &&(\mathbf{x}_i(s) - (\alpha \mathbf{x}_{i-1}(s) + (1-\alpha) \mathbf{x}_{1}(s))).
\end{eqnarray}

Let \(\mathbf{K}(s) = C_f \begin{pmatrix} k_e & k_{\theta} + k_w s \end{pmatrix}\). The components of the vectors \((\mathbf{x}_i - \mathbf{x}_{i-1})\) and \((\mathbf{x}_i - \mathbf{x}_{1})\) are the lateral and heading errors relative to the preceding and lead vehicles, respectively. Note that \(e_{lat, i}\) is the first component of \(\mathbf{x}_i\) and represents the lateral error from the straight line (\(R = \infty\)) the platoon is expected to follow. Similarly, \(\tilde{\theta}_i\) is the heading error relative to the straight line the \(i^{th}\) ACV is expected to track. However, the available measurements are relative errors, hence the controller's structure. The structure of \(\mathbf{K}(s)\) reflects the absence of lateral error rate information due to the noisy nature of lateral velocity information, which is unsuitable for feedback.

Using this control law, the error propagation equation is given by:
\[ (\mathbf{M}_0(s) + \mathbf{B}\mathbf{K}(s)) \mathbf{x}_i(s) = \mathbf{B}\mathbf{K}(s) (\alpha \mathbf{x}_{i-1}(s) + (1-\alpha) \mathbf{x}_{1}(s)). \]

We can rewrite the above equation as follows:
\begin{equation}
    \mathbf{x}_i(s) = \mathbf{G}(s) (\alpha \mathbf{x}_{i-1}(s) + (1-\alpha) \mathbf{x}_{1}(s)),
    \label{eq:x_i} 
\end{equation}
where
\begin{equation*}
    \mathbf{G}(s) = \left( ({\mathbf M}_0(s) + {\mathbf B}{\mathbf K}(s))^{-1} {\mathbf B} {\mathbf K}(s)\right).
\end{equation*}

Similarly, the equation for the \(i-1^{th}\) vehicle, as shown in \eqref{eq:x_i}, is:
\begin{equation}
    \mathbf{x}_{i-1}(s) = \mathbf{G}(s) (\alpha \mathbf{x}_{i-2}(s) + (1-\alpha) \mathbf{x}_{1}(s)).
    \label{eq:x_i-1} 
\end{equation}

Subtracting equation \eqref{eq:x_i-1} from \eqref{eq:x_i} results in the following:
\begin{equation}
    \mathbf{x}_i(s) - \mathbf{x}_{i-1}(s) = \alpha \mathbf{G}(s)(\mathbf{x}_{i-1}(s) - \mathbf{x}_{i-2}(s)).
    \label{eq:x_i-x_i-1}
\end{equation}

Since \(\mathbf{x}_2(s) = \mathbf{G}(s) \mathbf{x}_1(s)\), we can express the right-hand side of equation \eqref{eq:x_i-x_i-1} in terms of \(\mathbf{x}_1(s)\) as follows:
\begin{equation}
    \mathbf{x}_i(s) - \mathbf{x}_{i-1}(s) = (\alpha \mathbf{G}(s))^{i-2} (\mathbf{G}(s) - \mathbf{I}) \mathbf{x}_1(s).
\end{equation}

From \cite{vidyasagar1981input}, the two-norms of input and output are related through the ${\mathcal H}_{\infty}$ norm of the transfer function as:
\begin{equation*}
    \|\mathbf{x}_i(s) - \mathbf{x}_{i-1}(s)\|_{2} \leq \|\alpha \mathbf{G}(s)\|_{\infty}^{i-2} \cdot  \|\mathbf{G}(s) - \mathbf{I}\|_{\infty} \cdot \|\mathbf{x}_{1}(s)\|_{2}.
\end{equation*}

Let $\alpha$ be chosen so that 
$\rho:= \|\alpha \mathbf{G}(s)\|_{\infty} <1$. Then:
\begin{eqnarray}
    \nonumber
    \|{\bf x}_i\|_2 &=&  \|({\bf x}_i - {\bf x}_{i-1} ) + ( {\bf x}_{i-1} - {\bf x}_{i-2}) + \cdots +\\ 
    \nonumber
    &&({\bf x}_2 - {\bf x}_1) + {\bf x}_1)\|_2 \\  
    \nonumber
    &\le& \|{\bf x}_i - {\bf x}_{i-1}\|_2 + \cdots +
    \|{\bf x}_2 - {\bf x}_1\|_2 +  \|{\bf x}_1\|_2 \\
    \nonumber 
    &\le& ((\rho^{i-2} + \rho^{i-1} + \ldots + \rho +1)\cdot\\ 
    \nonumber 
    &&\|G(s)-I\|_{\infty} +1) \|{\bf x}_1\|_2 \\
    &\le& \underbrace{\frac{\|G(s)-I\|_{\infty}}{1-\rho}}_{M} \|{\bf x}_1(t)\|_2.
\end{eqnarray}

\noindent{\bf Remarks:}
\begin{enumerate}
    \item When $\alpha = 1$, only the preceding vehicle's information will be used in the feedback law. In that case $\rho = \|G(s)\|_{\infty}$ was shown to exceed $1$ at low frequencies in \cite{liu2024lateral}. 
    \item From the performance viewpoint, it is optimal to pick $\alpha = 0$ and it results in the smallest value of $M$. However, this is not advisable as the information about error of an ego ACV relative to the lead ACV, which requires connectivity and can suffer from significant delays in a longer convoy. 
\end{enumerate}






\section{RESULTS}
\label{sec:results}

\subsection{Vehicle Parameters and Feedback Gains}

The vehicle parameters used in this study, derived from experimental data for a Lincoln MKZ mid-size sedan, are consistent with prior work \cite{liu2024lateral}. Key parameters include the vehicle’s mass, front and rear axle loads, moment of inertia, and cornering stiffness for both front and rear axles. Additionally, the damping ratio (\(\zeta\)) and natural frequency (\(\omega_n\)) for steering actuation are specified as 0.4056 and 21.4813 rad/s, respectively, slightly differing from those used by Hassanain et al. \cite{hassanain2020string}.

The stabilizing feedback gains were computed across speeds of \(\{10, 20, 30, 40, 50, 60, 67\} \ \text{mph}\), with a controller gain vector \((k_e, k_{\theta}, k_{\omega}) = (0.06, 0.96, 0.08)\) chosen within the intersection of all stabilizing sets. Given that this selection lies within the bounded stabilizing region, these feedback gains ensure lateral stability, as demonstrated in prior validation \cite{liu2024lateral}.

\subsection{Simulation Results}

\begin{figure}[ht]
\centering
\includegraphics[width=0.3\textwidth]{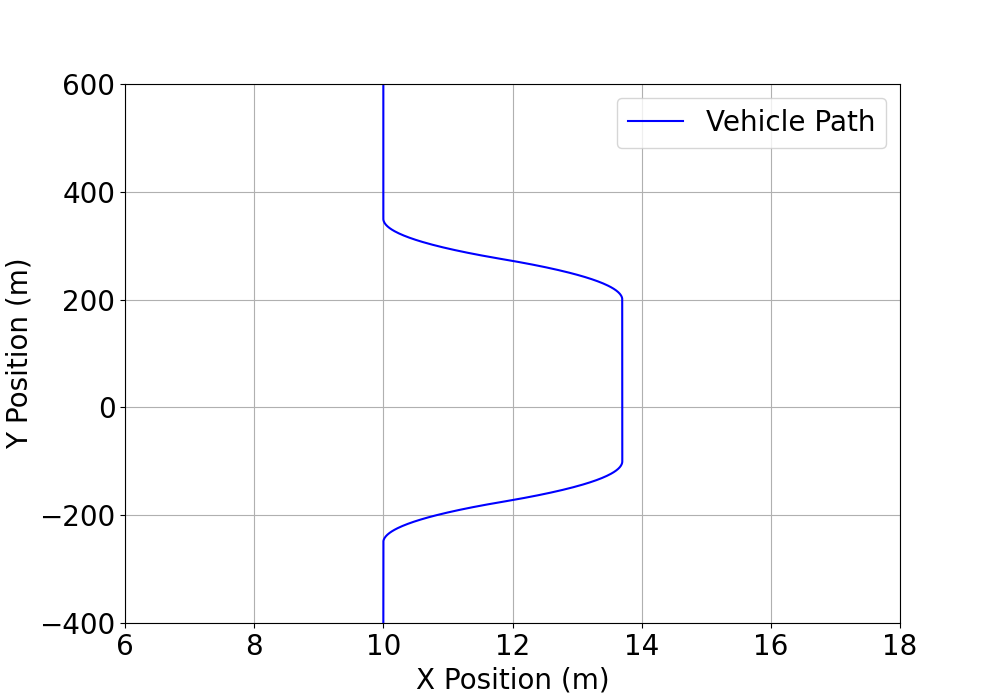}
\caption{Target trajectory for the lead vehicle.}
\label{fig:Path}
\end{figure}

Fig. \ref{fig:Path} illustrates a nominal double lane-change trajectory over a 1 km road segment, serving as the target path for the lead ACV. In this scenario, the lead ACV executes a rightward maneuver into the adjacent lane, followed by a leftward maneuver to return to the original lane. The second ACV, acting as the first follower in the platoon, uses sampled trajectory data from the lead ACV and applies the control law developed in this work. The remaining vehicles in the platoon access sampled trajectory data from both the lead ACV and their respective preceding ACVs, employing the lateral controller designed in the previous section.

\subsubsection{\bf Lateral controller performance}
We evaluate the performance of the proposed lateral control scheme based on separate ELC trajectories. The feedback and feedforward controllers, described in Section \ref{sec:design}, are equally weighted with \(\alpha = 0.5\).

\begin{figure}[ht]
    \centering
    \begin{subfigure}[b]{0.45\linewidth}
        \centering
        \includegraphics[width=\linewidth]{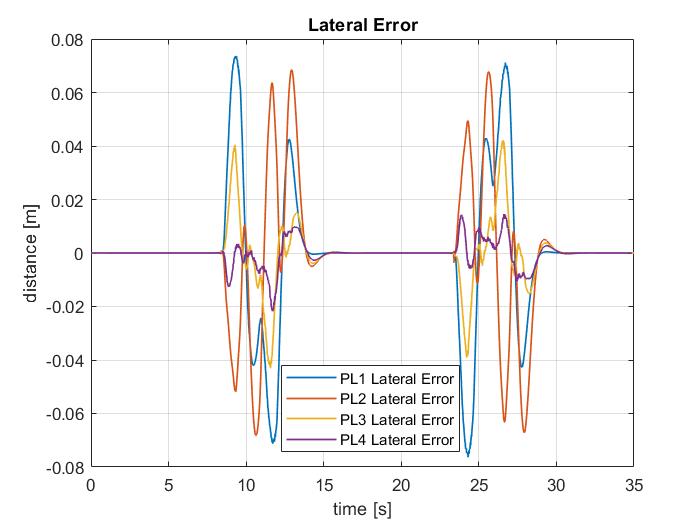}
        \caption{Lateral errors}
        \label{fig:LateralErrorSepAllHeavy}
    \end{subfigure}
    \hfill
    \begin{subfigure}[b]{0.45\linewidth}
        \centering
        \includegraphics[width=\linewidth]{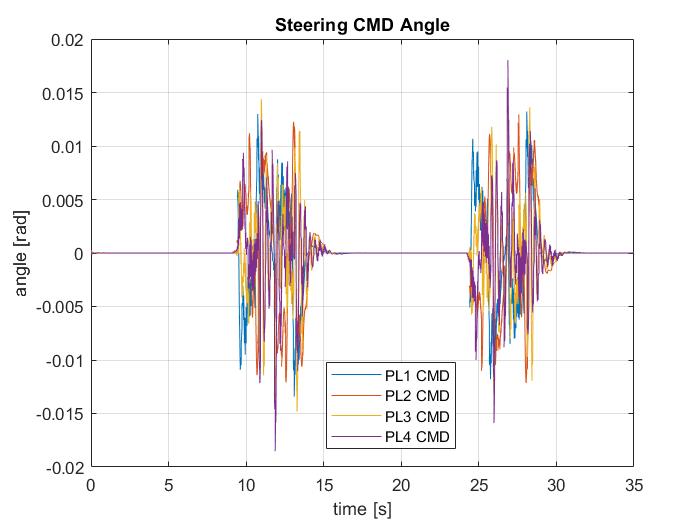}
        \caption{Steering command angles}
        \label{fig:FullSteeringCMDSep}
    \end{subfigure}
    \caption{Comparison of (a) lateral errors and (b) steering command angles of vehicles following the weighted separate ELC trajectories.}
    \label{fig:CombinedLateralErrorSteeringCMD}
\end{figure}

Fig. \ref{fig:LateralErrorSepAllHeavy} shows the lateral errors of the vehicles following the separate ELC trajectories. The analysis indicates that the lateral errors are well-contained, with the maximum error peaking at approximately \(8 \ \text{cm}\). This peak error occurs at the start and end of the turn, which are critical points in the maneuver. The overall trend shows a decrease in the maximum lateral error, confirming the effectiveness of the control scheme in maintaining stability and preventing lateral string instability.

Fig. \ref{fig:FullSteeringCMDSep} illustrates the steering command angles for the vehicles during the double lane change. The data shows the commanded steering angles over time for each vehicle in the platoon. The results indicate that the steering commands are within acceptable limits, contributing to the overall stability and performance of the lateral control system. 

\subsubsection{\bf Analysis of steering command angles for a bigger platoon}
In this analysis, we examine the steering command angles of vehicles in the platoon during the ELC maneuver, focusing on how these commands scale with the vehicle index for both proposed trajectory approaches. To conduct this analysis, a simulation of a 10-ACV platoon performing a double lane change, as depicted in Fig. \ref{fig:Path}, was conducted. Additionally, the performance of the proposed lateral control scheme is compared with the lateral control method proposed in \cite{liu2024lateral}.

\begin{figure}[ht]
    \centering
    \begin{subfigure}[b]{0.45\linewidth}
        \centering
        \includegraphics[width=\linewidth]{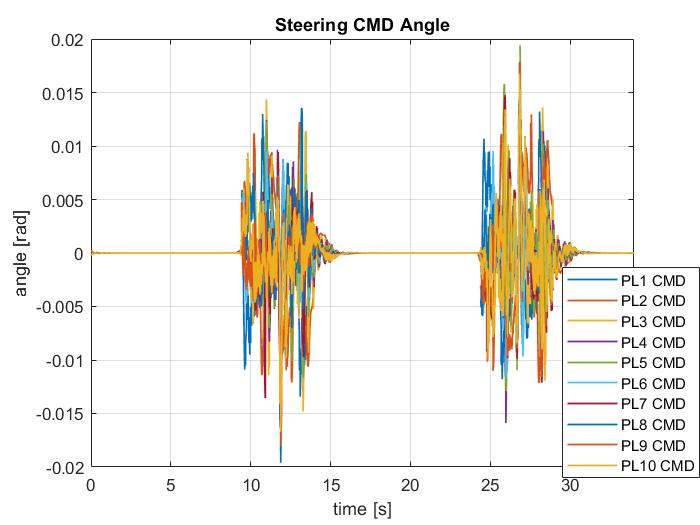}
        \caption{Separate ELC trajectories.}
        \label{fig:FullSteeringCMDSep10vehicles}
    \end{subfigure}
    \hfill
    \begin{subfigure}[b]{0.45\linewidth}
        \centering
        \includegraphics[width=\linewidth]{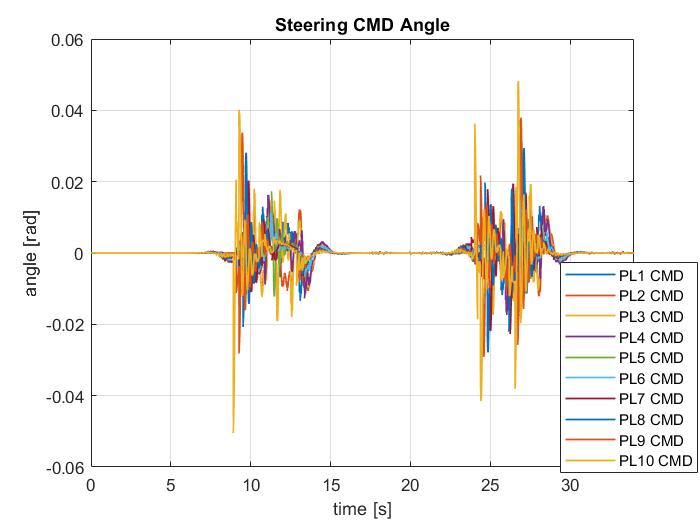}
        \caption{Composite ELC trajectory.}
        \label{fig:FullSteeringCMDCombine10vehicles}
    \end{subfigure}
    \caption{Comparison of steering command angles for a 10-ACV platoon following separate and composite \cite{liu2024lateral} ELC trajectory methods shown in (a) and (b), respectively.}
    \label{fig:CombinedSteeringCMD}
\end{figure}

Fig. \ref{fig:FullSteeringCMDSep10vehicles} depicts the steering command angles for a 10-ACV platoon following the weighted separate ELC trajectories. Fig. \ref{fig:FullSteeringCMDCombine10vehicles} illustrates the steering command angles of vehicles in the same 10-ACV platoon, but following the composite ELC trajectory derived from the method in \cite{liu2024lateral}. Upon comparison, it is evident that the maximum steering command angle is significantly lower for the lateral control scheme proposed in this paper compared to the composite ELC trajectory case. Additionally, unlike the composite trajectory approach, the proposed separate trajectories method do not show a monotonically increasing trend in steering command angles with the vehicle index. The maximum steering angle command remains nearly constant for both the 4-ACV platoon (as shown in Fig. \ref{fig:FullSteeringCMDSep}) and the 10-ACV platoon (as shown in Fig. \ref{fig:FullSteeringCMDSep10vehicles}).

\section{CONCLUSIONS}
\label{sec:conclusions}
This paper presents a comprehensive study on the lateral control of Autonomous and Connected Vehicles (ACVs) in a platoon executing an Emergency Lane Change (ELC) maneuver. We propose a lateral control framework that leverages communicated position data from both the lead and preceding vehicles to construct two separate ELC trajectories based on limited GPS data. This framework integrates both feedforward and feedback controllers to track real-time trajectories by computing error signals. Additionally, we establish the lateral string stability of a platoon executing a straight-line maneuver by utilizing information from both the lead and preceding vehicles, addressing the limitations of previous works that demonstrated instability with predecessor-only information. Simulation results indicate that the proposed control scheme effectively manages the dynamic requirements of the ELC maneuver, ensuring the safe and stable operation of ACVs in a platoon configuration, as lateral errors are maintained within acceptable limits and show a decreasing trend. Analysis of steering command angles across different platoon sizes demonstrates that the proposed lateral control method does not exhibit a monotonically increasing trend in steering command angles with platoon size, highlighting its scalability and robustness.

\bibliographystyle{IEEEtran}
\bibliography{References}

\end{document}